\documentclass[11pt,english,letterpaper]{revtex4}
\usepackage[T1]{fontenc}
\usepackage[latin1]{inputenc}
\usepackage{geometry}
\usepackage{graphicx}
\usepackage{amssymb}
\makeatletter

\usepackage{babel}
\makeatother
\usepackage{colordvi}

\DeclareMathAlphabet{\mathsfsl}{OT1}{cmss}{m}{sl}

\renewcommand{\vec}[1]{\mbox{\boldmath$#1$}}

\newcommand{\xh}{\hat{\vec{x}}}
\newcommand{\yh}{\hat{\vec{y}}}
\newcommand{\zh}{\hat{\vec{z}}}

\newcommand{\nab}{\vec{\nabla}}

\newcommand{\rb}[1]{\left( #1 \right)}
\renewcommand{\sb}[1]{\left[ #1 \right]}

\newcommand{\B}{\vec{B}}

\newcommand{\V}{\vec{V}}

\begin{document}
%**********************************************************************************

\title{Analytical Model of Fast Magnetic Reconnection with a Large Guide Field}
\author{Andrei N. Simakov$^{1,2,\ast}$, L. Chac\'{o}n$^3$, and A. Zocco$^4$}
\affiliation{$^1$Theoretical Division, Los Alamos National Laboratory, Los Alamos, NM 87545, USA \\ $^2$Max-Planck-Institut f\"ur Plasmaphysik, 17491 Greifswald, Germany \\ $^3$Fusion Energy Division, Oak Ridge National Laboratory, Oak Ridge, TN 37830, USA \\ $^4$Culham Centre for Fusion Energy, Abingdon, Oxon, OX14 3DB, UK}

\email{simakov@lanl.gov}

%**********************************************************************************
\begin{abstract}
Analytical theory of fast magnetic reconnection with a large guide field is presented for the first time. We confirm that two distinct reconnection regimes are possible depending on whether the diffusion region thickness $\delta$ is larger or smaller than the sound gyroradius $\rho_s$. The reconnection is slow or Sweet-Parker-like for $\delta \gtrsim \rho_s$, and fast otherwise. In the fast regime, however, we find that ion viscosity $\mu$ plays a critical role. In particular, for $\delta < \rho_s$ the diffusion region thickness is proportional to $Ha^{-1}$ with $Ha \propto 1/\sqrt{\eta \mu}$ the Hartmann number, and the reconnection rate is proportional to $Pr^{-1/2}$ with $Pr = \mu/\eta$ the Prandtl number and $\eta$ the resistivity. If the perpendicular ion viscosity is employed for $\mu$ the reconnection rate becomes independent of both plasma $\beta$ and collision frequencies and therefore potentially fast.
\end{abstract}
\maketitle

%**********************************************************************************
Fast magnetic reconnection phenomena observed in space and laboratory plasmas indicate that nonlinear reconnection rates are independent of collisional dissipation coefficients. Such dissipation independent rates have been reproduced computationally for reconnection with \cite{kleva_95,ma_96,fitzpatrick_04,huba_05,schmidt_09} and without \cite{birn_01} a guide magnetic field. They have also been predicted analytically in electron \cite{chacon_07} and Hall \cite{simakov_08,simakov_09,malyshkin_08} magnetohydrodynamics (MHD) models {\it without a guide field}. However, no analytical model for reconnection {\it with a guide field} is currently available. Since a guide field is normally present and sometimes significantly exceeds the reconnecting in-plane magnetic field components, as in magnetic fusion experiments, a nonlinear analytical model of fast reconnection with a guide field is highly desirable.

Extended MHD simulations of guide-field reconnection have demonstrated that fast reconnection regimes \cite{kleva_95,ma_96,fitzpatrick_04,huba_05,schmidt_09} can be enabled by the electron pressure gradient in the Ohm's law \cite{aydemir_92} and occur when the sound gyroradius $\rho_s$, defined as the ion gyroradius with electron temperature, exceeds the diffusion region thickness \cite{kleva_95,schmidt_09,cassak_07}. The latter conclusion was recently confirmed experimentally \cite{egedal_07}. However, despite significant theoretical progress, a number of fundamental questions remains unanswered. In particular, the scaling of the nonlinear reconnection rate with $\rho_s$ is not unique and appears problem dependent: Kleva {\it et al.} \cite{kleva_95} found $E_z \propto \rho_s^0$ for magnetic flux bundle coalescence; Bhattacharjee {\it et al.} \cite{bhattacharjee_05} observed $E_z \propto \rho_s$ for tearing modes; Fitzpatrick \cite{fitzpatrick_04} concluded that $E_z \propto \rho_s^{3/2}$ for the Taylor problem; while Schmidt {\it et al.} \cite{schmidt_09} saw $E_z \propto \rho_s^{\alpha}$ for tearing modes and magnetic island coalescence, where $\alpha$ decreased from 1 to 0 as $\rho_s$ increased. Formation of a thin current sub-layer inside the current sheet layer, which results in quasi-explosive reconnection during early nonlinear stages \cite{bhattacharjee_05,ottaviani_93,ottaviani_95,cafaro_98,ramos_02}, and its nonlinear saturation are also poorly understood. Simulations show that the total layer thickness remains of order the electron skin depth $d_e$ \cite{ottaviani_95,chacon_08_1}, whereas the sub-layer thins with time without a limit. It has been speculated \cite{ottaviani_95,cafaro_98,ramos_02} (but not demonstrated) that such thinning in unphysical and will eventually be arrested by three-dimensional nonlinearities or electron viscosity $\mu_e$. Ion viscosity $\mu$ is usually believed to play no significant role in the large-guide-field fast reconnection process (e.g., see Refs.~\cite{kleva_95,fitzpatrick_04,schmidt_09}).

Here we propose, for the first time, a simple {\it nonlinear analytical} model of magnetic reconnection in extended MHD with a {\it large guide field}. We consider a quasi-steady state diffusion region in two dimensions and evaluate its aspect ratio and the corresponding reconnection rate. We take into account plasma resistivity $\eta$, ion viscosity $\mu$, as well as finite sound gyroradius $\rho_s$ effects. Similar to the zero-guide-field case \cite{chacon_07,simakov_08,simakov_09,malyshkin_08}, we find that two reconnection regimes are possible: a fast and a slow, Sweet-Parker-like \cite{sweet_parker}. But, unlike the zero-guide-field case, where the reconnection rate in the fast regime is not explicitly dependent on dissipation coefficients, here we find it proportional to $Pr^{-1/2}$ with $Pr = \mu/\eta$ the magnetic Prandtl number. Assuming \cite{biskamp_book} that $\mu$ corresponds to the perpendicular ion viscosity \cite{braginskii} gives $Pr^{-1/2} \sim \beta^{-1/2} (m_e/m_i)^{1/4}(T_i/T_e)^{1/4}$, which predicts the reconnection rate not to be explicitly dependent on collision frequencies. Here $\beta \ll 1$ is the ratio of plasma pressure to magnetic pressure, and $m_e$, $m_i$, $T_e$, and $T_i$ are electron and ion masses and temperatures, respectively. Therefore, contrary to previous beliefs \cite{kleva_95,fitzpatrick_04,schmidt_09}, ion viscosity plays a fundamental role in the fast reconnection process with a guide field. A transition between the two regimes occurs when the current layer thickness becomes comparable with $\rho_s$, in agreement with previous findings \cite{kleva_95,schmidt_09,cassak_07,egedal_07}. The model gives an explicit dependence of $E_z$ on $\rho_s$ and suggests why different scalings can be observed for different reconnecting systems. It also obtains a nonlinear threshold for the current sheet thinning \cite{bhattacharjee_05,ottaviani_93,ottaviani_95,cafaro_98,ramos_02} and demonstrates how it is arrested by dissipation processes.

{\it Two-field fluid model.} -- We consider plasma in a strong, straight, homogeneous magnetic field $\B_0 = B_0 \zh$, $\zh \equiv \nab z$, and employ the two-dimensional ($\partial/\partial z \equiv 0$), low-$\beta$, cold-ion fluid model of Refs.~\cite{kleva_95,schmidt_09,cafaro_98,bhattacharjee_05,biskamp_book,schep_94} for magnetic reconnection in the $x - y$ plane. For simplicity, we neglect in this work finite electron inertia $d_e$ effects. These effects are important and will be discussed in future work. Normalizing to the Alfv\'en speed $V_A \equiv B_0/\sqrt{4 \pi n_0 m_i}$, with $n_0$ the homogeneous {\it equilibrium} plasma density, and an arbitrary length $L$ gives
\begin{eqnarray}
\rb{\frac{\partial}{\partial t} + \vec{V} \cdot \nab} \varpi = \vec{B} \cdot \nab (\nabla^2 \psi) + \mu \nabla^2 \varpi, \label{vort_eqn} \\ \frac{\partial \vec{B}}{\partial t} + \nab \times [\vec{B} \times (\vec{V} - \rho_s^2 \zh \times \nab \varpi)] = - \eta \nab \times (\nab \times \vec{B}). \label{B_eqn}
\end{eqnarray}
Here, $\V \equiv \zh \times \nab \varphi$ is the ion flow velocity with $\varphi$ the electrostatic potential, $\varpi \equiv \nab^2 \varphi$ is the flow vorticity, and $\B \equiv \zh \times \nab \psi$ is the magnetic field in the $x - y$ plane. The sound gyroradius is $\rho_s \equiv \sqrt{T_e/m_i}/(\Omega_i L)$ with $\Omega_i \equiv e B_0/(m_i c)$ the ion gyrofrequency, $c$ the speed of light, and $e$ the proton electric charge. Finally, $\eta$ and $\mu$ are normalized resistivity and ion viscosity, respectively. In numerical simulations, Eq.~(\ref{B_eqn}) is often replaced with a single scalar equation for $\psi$. However, for our purposes \cite{chacon_07,simakov_08,simakov_09,chacon_08_1}, it is convenient to consider the vector form of the equation.

\begin{figure}[h]
\begin{center}
\includegraphics[width=3.0in,keepaspectratio]{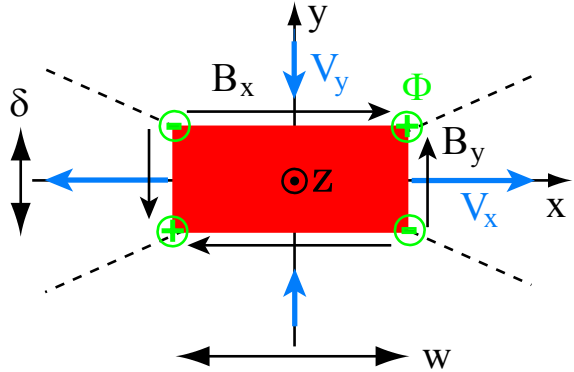}
\end{center}
\caption{(color online). Diffusion region geometry.}
\label{geometry}
\end{figure}

We concentrate on the diffusion region and do not concern ourselves with a particular physical system supplying magnetic flux. A closed dynamical description can be obtained by coupling our microscopic diffusion region description with a suitable macroscopic driver, as in Ref.~\cite{simakov_06}. As shown in Fig.~\ref{geometry}, we assume a rectangular diffusion region of (normalized) dimensions $\delta$ and $w$ with plasma entering and exiting along the $\yh \equiv \nab y$ and $\xh \equiv \nab x$ directions, respectively. We define the discrete upstream and downstream magnetic field variables $B_x \equiv \xh \cdot \B(0,\delta/2)$ and $B_y \equiv \yh \cdot \B(w/2,0)$, respectively, and the discrete flow stream function $\Phi \equiv - \varphi(w/2,\delta/2)$. Then, the inflow and outflow velocities are given by $V_y = - 2 \Phi/w$ and $V_x = 2 \Phi/\delta$, respectively. Following Refs.~\cite{chacon_07,simakov_08,simakov_09,chacon_08_1} we next discretize Eq.~(\ref{vort_eqn}) at $(x,y)=(w/2,\delta/2)$ and the $\xh$ and $\yh$ components of Eq.~(\ref{B_eqn}) at $(x,y)=(0,\delta/2)$ and $(w/2,0)$, respectively. Time-derivative terms are normally found small in nonlinearly saturated states, e.g. at and around the time of the local maximum of the reconnection rate \cite{simakov_06}. A steady-state analysis of the diffusion region (but not the full domain) is appropriate in such situations. Neglecting time derivatives and numerical factors of order unity and following a similar procedure as in Refs.~\cite{chacon_07,simakov_08,simakov_09,chacon_08_1} results in the following equations for $\Phi$, $B_x$, and $B_y$:
\begin{eqnarray}
\frac{\Phi^2}{\delta w} \rb{\frac{1}{\delta^2} - \frac{1}{w^2}} + \rb{\frac{B_x}{w} + \frac{B_y}{\delta}} \rb{\frac{B_y}{w} - \frac{B_x}{\delta}} = - \mu \Phi \rb{\frac{1}{\delta^2} + \frac{1}{w^2}}^2, \label{eq1} \\ - \frac{\Phi B_x}{\delta w} \sb{1 + \rho_s^2 \rb{\frac{1}{\delta^2} + \frac{1}{w^2}}} = \eta \rb{\frac{B_y}{\delta w} - \frac{B_x}{\delta^2}}, \label{eq2} \\ \frac{\Phi B_y}{\delta w} \sb{1 + \rho_s^2 \rb{\frac{1}{\delta^2} + \frac{1}{w^2}}} = \eta \rb{\frac{B_x}{\delta w} - \frac{B_y}{w^2}}. \label{eq3}
\end{eqnarray}
Equations (\ref{eq1}) -- (\ref{eq3}) are invariant under the plasma flow reversal, i.e., under the substitution $(B_x,B_y,\Phi,\delta) \leftrightarrow (B_y,B_x,-\Phi,w)$, as expected \cite{chacon_07,chacon_08_1}. These three equations contain five unknowns: $\delta$, $w$, $B_x$, $B_y$, $\Phi$. It is therefore necessary to consider two independent parameters. We have found it convenient to choose $w$ and $B_x$ as parameters, as they can be found via coupling with a macroscopic driver \cite{simakov_06}.

{\it Solution of the discrete equations.} -- An important characteristic of the diffusion region is its aspect ratio $\xi \equiv \delta/w$. The reconnection rate $E_z = -\eta J_z = \eta (B_x/\delta - B_y/w)$ can be conveniently expressed in terms of $\xi$ as
\begin{equation}
E_{z \ast} \equiv \frac{E_z}{B_x^2} = \frac{\sqrt{2} (1 - \xi^2)}{S_{\eta} \xi}, \label{Ez}
\end{equation}
where $S_{\eta} \equiv \sqrt{2} B_x w/\eta$ is the resistive Lundquist number. It is clear from Eq.~(\ref{Ez}) that, for given $B_x$ and $w$, large reconnection rates preferentially occur for $\xi \ll 1$. We will concentrate next on the $\xi < 1$ limit by approximating $1 + \xi^2 \approx 1 - \xi^2 \approx 1$ in Eq.~(\ref{Ez}) and elsewhere. Then, the reconnection rate becomes $E_{z \ast} \approx \sqrt{2}/(S_{\eta} \xi)$. An equation for $\xi$, or equivalently $\delta$, can be obtained from Eqs.~(\ref{eq1}) -- (\ref{eq3}). Introducing a dimensionless quantity $\hat{\delta} \equiv \delta/\rho_s$, which characterizes importance of plasma compressibility, and eliminating variables results in the following equation for $\xi$:
\begin{equation}
\frac{1}{S_{\eta}^2 \xi^4} + \frac{1 + \hat{\delta}^2}{S_{\eta} S_{\mu} \xi^4 \hat{\delta^2}} = \frac{(1 + \hat{\delta}^2)^2}{\hat{\delta}^4}, \label{master1}
\end{equation}
where $S_{\mu} \equiv \sqrt{2} B_x w/\mu$ is the viscous Lundquist number.

To solve Eq.~(\ref{master1}) for $\xi$, or equivalently $\delta = \xi w$, it is convenient to introduce parameters $q \equiv Pr/(Pr + 1) \leq 1$, with $Pr \equiv \mu/\eta$ the magnetic Prandtl number; and $\hat{\delta}_{SP} \equiv \delta_{SP}/\rho_s$, with $\delta_{SP} \equiv (w/\sqrt{S_{\eta}}) (1 + Pr)^{1/4}$ the Sweet-Parker length scale. Then, Eq.~(\ref{master1}) becomes
\begin{equation}
\hat{\delta}^2 (\hat{\delta}^2 + 1)^2 - \hat{\delta}_{SP}^4 (\hat{\delta}^2 + q) = 0, \label{master2}
\end{equation}
so that
\begin{equation}
\delta \approx \left\{ \begin{array}{cc} \delta_{SP}, & \delta_{SP} \gg \rho_s \\ \sqrt{q} \delta_{SP}^2/\rho_s = w^2/(\rho_s Ha), & \delta_{SP} \ll \rho_s \end{array} \right. \label{asymptotics_delta},
\end{equation}
where $Ha = \sqrt{S_{\eta} S_{\mu}}$ is the Hartmann number. The corresponding reconnection rates are
\begin{equation}
E_{z \ast} \approx \left\{ \begin{array}{cc} \sqrt{2/S_{\eta}} (Pr + 1)^{-1/4}, & \delta_{SP} \gg \rho_s \\ \sqrt{2/Pr} (\rho_s/w), & \delta_{SP} \ll \rho_s \end{array} \right. \label{asymptotics_Ez}.
\end{equation}

\begin{figure}[b]
\begin{center}
\includegraphics[width=3.5in,keepaspectratio]{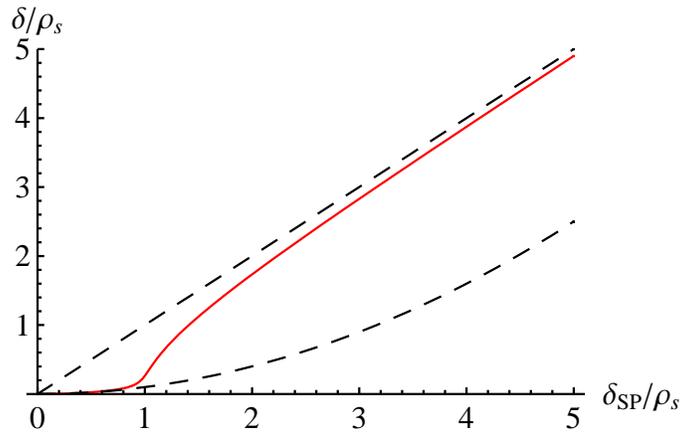}
\end{center}
\caption{(color online). Numerical solution of Eq.~(\ref{master2}) with $q = 0.01$ (solid line). Asymptotics (\ref{asymptotics_delta}) are shown with dashed lines.}
\label{fig2}
\end{figure}

Consequently, two reconnection regimes are possible. When the diffusion region thickness exceeds $\rho_s$ the reconnection is Sweet-Parker like \cite{sweet_parker} and therefore slow. In the opposite limit the diffusion region thickness is proportional to $Ha^{-1}$ and the reconnection rate to $Pr^{-1/2} = \sqrt{\eta/\mu}$. Therefore, both ion viscosity and resistivity are essential ingredients. However \cite{biskamp_book}, if $\mu$ corresponds to the perpendicular ion viscosity, then $Pr^{-1/2} \sim \beta^{-1/2} (m_e/m_i)^{1/4}(T_i/T_e)^{1/4}$ and the reconnection rate is not explicitly dependent on collision frequencies. Hence, it is potentially fast. Since $\rho_s = \sqrt{\beta/2} d_i$ with $d_i = (c/L) \sqrt{m_i/(4 \pi n e^2)}$ the normalized ion inertial length scale, the reconnection rate $E_{z \ast} = (2/Pr)^{1/2} (\rho_s/w)$ in the fast reconnection regime is independent of $\beta \ll 1$. A transition between the two regimes occurs at $\delta \sim \delta_{SP} \sim \rho_s$, as expected \cite{kleva_95,schmidt_09,cassak_07,egedal_07}. These conclusions are confirmed by solving Eq.~(\ref{master2}) numerically. The solution for $q = 0.01$ is shown in Fig.~\ref{fig2} with a solid line. The asymptotics (\ref{asymptotics_delta}) are shown with dashed lines.

It follows from Eqs.~(\ref{asymptotics_delta}) and (\ref{asymptotics_Ez}) that the fast reconnection regime does not exist in steady state for $\mu = 0$. This corresponds to the absence of a steady-state solution and the indefinite current sub-layer thinning observed in Refs.~\cite{bhattacharjee_05,ottaviani_93,ottaviani_95,cafaro_98,ramos_02}. In fact, Eq.~(\ref{master2}) predicts that a steady-state solution in this case only exists for $\delta_{SP} \geq \rho_s$. Finite ion viscosity regularizes the solution for $\delta, \delta_{SP} \lesssim \rho$. Similarly, Eqs.~(\ref{asymptotics_delta}) and (\ref{asymptotics_Ez}) predict that the fast reconnection regime does not exist when $\rho_s = 0$ for arbitrary $\mu$, as expected.

{\it Fundamental role of ion viscosity.} -- The relevance of $\mu$ in the fast reconnection regime with a large guide field can be understood directly from Eqs.~(\ref{vort_eqn}) and (\ref{B_eqn}). Consider a quasi-steady state diffusion layer and assume $\xi \ll 1$, so that according to Eqs.~(\ref{eq2}) and (\ref{eq3}) $B_y/B_x = \xi \ll 1$. Equation (\ref{vort_eqn}) requires the term $\B \cdot \nab (\nabla^2 \psi)$ to be balanced by either $\V \cdot \nab \varpi$ or $\mu \nabla^2 \varpi$. Therefore, viscosity is important if $\mu \nabla^2 \sim \mu/\delta^2 > \V \cdot \nab \sim \Phi/(\delta w)$ and unimportant otherwise. The magnitude of $\Phi$ can be estimated from Eq.~(\ref{B_eqn}) by discretizing its $x$ component at $(x,y) = (0,\delta/2)$:
\begin{equation}
\frac{B_x \Phi}{\delta w} \rb{1 + \frac{\rho_s^2}{\delta^2}} \sim \frac{\eta B_x}{\delta^2}.
\end{equation}
Then,
\begin{equation}
\frac{\Phi}{\delta w} \sim \left\{ \begin{array}{cc} \eta/\delta^2, & \delta > \rho_s \\ \eta /\rho_s^2, & \delta < \rho_s \end{array} \right. . \label{estimate_Phi}
\end{equation}
Since $\eta$ and $\mu$ have the same scaling with collisionality when $\mu$ represents perpendicular ion viscosity, and since $\delta$ decreases with collisionality, for $\delta < \rho_s$ the viscous term in the vorticity equation will dominate over the advective term for small collision frequencies and $\mu$ will play a fundamental role. On the other hand, for $\delta > \rho_s$ both terms are comparable and $\mu$ does not influence reconnection fundamentally. Employing result (\ref{estimate_Phi}) in Eq.~(\ref{vort_eqn}) discretized at $(x,y) = (\delta/2,w/2)$,
\begin{equation}
\frac{\Phi^2}{\delta^3 w} + \frac{\mu \Phi}{\delta^4} \sim \frac{B_x^2}{\delta w},
\end{equation}
gives result (\ref{asymptotics_delta}) for $\delta$ and consequently result (\ref{asymptotics_Ez}) for $E_{z \ast}$, as expected.

{\it Comparison with previous numerical studies.} -- The results obtained herein agree well with the available numerical simulations of guide field reconnection. Specifically, Kleva {\it et al.} \cite{kleva_95} employed Eqs.~(\ref{vort_eqn}) and (\ref{B_eqn}) with $\mu = \eta$ (i.e., $Pr = 1$) to study coalescence of two magnetic flux bundles. The work varied $\rho_s$ and $\eta$ in the fast reconnection regime ($\delta < \rho_s$) by as much as a factor of four each and found $w \propto \rho_s$ and $E_z \propto \eta^0, \rho_s^0$, consistent with Eq.~(\ref{asymptotics_Ez}). In addition, it was observed that $\delta \propto \eta$, in agreement with Eq.~(\ref{asymptotics_delta}) when $\mu=\eta$.

Schmidt {\it et al.} \cite{schmidt_09} also employed Eqs.~(\ref{vort_eqn}) and (\ref{B_eqn}) with $\mu = \eta$ (again, $Pr = 1$), but studied reconnection driven by a tearing mode and by a magnetic island coalescence instability. In both cases, the reconnection rate in the fast reconnection regime ($\delta < \rho_s$) was again found to be independent of $\eta$, $E_z \propto \eta^0$. Moreover, it was observed that $E_z \propto \rho_s^{\alpha}$ with $\alpha \sim 1$ for $\delta \sim \delta_{SP} \sim \rho_s$ (i.e., at the slow-to-fast transition) and $\alpha \sim 0$ for $\rho_s \gg \delta_{SP}$. Both scalings are also consistent with Eq.~(\ref{asymptotics_Ez}) when one considers that, at the transition, $w$ is still determined by the Sweet-Parker dynamics and is therefore independent of $\rho_s$ (i.e., $w \propto \rho_s^0$; hence $E_z \propto \rho_s/w \propto \rho_s$), while $w \propto \rho_s$ when $\delta$ is deep in the $\rho_s$ sub-layer \cite{kleva_95}, resulting in $E_z \propto \rho_s/w \propto \rho_s^0$.

One set of results apparently at odds with our predictions is documented in Ref.~\cite{fitzpatrick_04}, where a system of four nonlinear equations for $\psi$, $\varpi$, and the $\zh$ components of the perturbed magnetic field and ion flow velocity is employed that is valid for both $\beta \ll 1$ and $\beta \gg 1$ regimes. When solving this system in the nonlinear reconnection regime with a large guide field, it was found that $E_z \propto \eta^0 \rho_s^{3/2}$ with fixed $\mu$, which appears to disagree with our findings (\ref{asymptotics_delta}) and (\ref{asymptotics_Ez}). However, no data is provided for $w$ in the reference. Furthermore, these four equations can only be reduced to our Eqs.~(\ref{vort_eqn}) and (\ref{B_eqn}) when $\beta \ll 1$, $\kappa = \mu \ll \beta \eta$, and $\nu \equiv 0$, with $\kappa$ the plasma heat conductivity and $\nu$ the hyper-resistivity as defined in Ref.~\cite{fitzpatrick_04}. Since Ref.~\cite{fitzpatrick_04} employed $\kappa = \mu \sim \beta \eta$ and $\nu = 2.5 \times 10^{-9} > 0$, the results therein cannot yet be compared with our theory.

In conclusion, we have presented a simple nonlinear analytical model of magnetic reconnection with a large guide field. We found that two reconnection regimes are possible depending on whether the diffusion region (current layer) thickness $\delta$ is larger or smaller than the sound gyroradius $\rho_s$. For $\delta > \rho_s$ the reconnection is slow and is described by the standard Sweet-Parker expressions modified to account for ion viscosity $\mu$. For $\delta < \rho_s$ the character of reconnection changes. Instead of scaling with $S_{\eta}^{-1/2}$ the diffusion region thickness becomes proportional to $Ha^{-1}$ with $Ha = \sqrt{S_{\eta} S_{\mu}}$ the Hartmann number; and the reconnection rate becomes proportional to $Pr^{-1/2}$ with $Pr = \mu/\eta$ the Prandtl number. Assuming that $\mu$ describes the perpendicular ion viscosity results in a reconnection rate that is independent of collision frequencies and plasma $\beta$ and is therefore potentially fast. The transition between the two regimes occurs at $\delta \sim \delta_{SP} \sim \rho_s$. Therefore, contrary to the common belief \cite{kleva_95,fitzpatrick_04,schmidt_09}, ion viscosity is found to play a fundamental role in large-guide-field reconnection. In particular, a steady-state fast branch does not exist for $\mu = 0$, resulting in indefinite thinning of $\delta$ with time \cite{bhattacharjee_05,ottaviani_93,ottaviani_95,cafaro_98,ramos_02} when $\delta < \rho_s$.

Given the demonstrated relevance of ion viscosity for the low-$\beta$ Hall MHD reconnection, it seems clear that the simple resistive and viscous closures commonly used in the literature for Eqs.~(\ref{vort_eqn}) and (\ref{B_eqn}) are likely to be inadequate for accurately describing weakly collisional magnetospheric and magnetic fusion plasmas. It follows that more accurate models for collisionless electron and ion viscosities (e.g., gyroviscosities \cite{braginskii}) must be employed to adequately explain experimental observations. The consideration of such closures in our theoretical framework will be the subject of future work.

Acknowledgement. We acknowledge helpful discussions with D. Grasso and D. Borgogno. This work was performed at Max-Planck-Institut f\"ur Plasmaphysik, Greifswald, Germany, at Los Alamos National Laboratory, operated by Los Alamos National Security LLC under contract DE-AC52-06NA25396, and at Oak Ridge National Laboratory, operated by UT-Battelle under contract DE-AC05-00OR22725.

%**********************************************************************************

%**********************************************************************************

\begin{thebibliography}{99}

\bibitem{kleva_95} R. G. Kleva, J. F. Drake, and F. L. Waelbroeck, Phys. Plasmas {\bf 2}, 23 (1995).

\bibitem{ma_96} Z. W. Ma and A. Bhattacharjee, Geophys. Res. Lett. {\bf 23}, 1673 (1996).

\bibitem{fitzpatrick_04} R. Fitzpatrick, Phys. Plasmas {\bf 11}, 3961 (2004).

\bibitem{huba_05} J. D. Huba, Phys. Plasmas {\bf 12}, 012322 (2005) and references therein.

\bibitem{schmidt_09} S. Schmidt, S. G\"unter, and K. Lackner, Phys. Plasmas {\bf 16}, 072302 (2009).

\bibitem{birn_01} J. Birn {\it et al.}, J. Geophys. Res. {\bf 106}, 3715 (2001) and references therein.

\bibitem{chacon_07} L. Chac\'on, A. N. Simakov, and A. Zocco, Phys. Rev. Lett. {\bf 99}, 235001 (2007).

\bibitem{simakov_08} A. N. Simakov and L. Chac\'on, Phys. Rev. Lett. {\bf 101}, 105003 (2008).

\bibitem{simakov_09} A. N. Simakov and L. Chac\'on, Phys. Plasmas {\bf 16}, 055701 (2009).

\bibitem{malyshkin_08} L. M. Malyshkin, Phys. Rev. Lett. {\bf 101}, 225001 (2008).

\bibitem{aydemir_92} A. Y. Aydemir, Phys. Fluids B {\bf 4}, 3469 (1992).

\bibitem{cassak_07} P. A. Cassak, J. F. Drake, and M. A. Shay, Phys. Plasmas {\bf 14}, 054502 (2007).

\bibitem{egedal_07} J. Egedal, W. Fox, N. Katz, M. Porkolab, K. Reim, and E. Zhang, Phys. Rev. Lett. {\bf 98}, 015003 (2007).

\bibitem{bhattacharjee_05} A. Bhattacharjee, K. Germaschewski, and C. S. Ng, Phys. Plasmas {\bf 12}, 042305 (2005).

\bibitem{ottaviani_93} M. Ottaviani and F. Porcelli, Phys. Rev. Lett. {\bf 71}, 3802 (1993).

\bibitem{ottaviani_95} M. Ottaviani and F. Porcelli, Phys. Plasmas {\bf 2}, 4104 (1995).

\bibitem{cafaro_98} E. Cafaro, D. Grasso, F. Pegoraro, F. Porcelli, and A. Saluzzi, Phys. Rev. Lett. {\bf 80}, 4430 (1998).

\bibitem{ramos_02} J. J. Ramos, F. Porcelli, and R. Ver\'astegui, Phys. Rev. Lett. {\bf 89}, 055002 (2002).

\bibitem{chacon_08_1} L. Chac\'on, A. N. Simakov, V. S. Lukin, and A. Zocco, Phys. Rev. Lett. {\bf 101}, 025003 (2008).

\bibitem{sweet_parker} P. A. Sweet, in {\it Electromagnetic Phenomena in Cosmic Physics}, ed. B. Lehnert (Cambridge Univ. Press, Cambridge, England, 1958); E. N. Parker, J. Geophys. Res. {\bf 62}, 509 (1957).

\bibitem{biskamp_book} D. Biskamp, {\it Magnetic Reconnection in Plasmas} (Cambridge Univ. Press, New York, 2000).

\bibitem{braginskii} S. I. Braginskii, in {\it Reviews of Plasma Physics}, edited by M. A. Leontovich (Consultants Bureau, New York, 1965), Vol. 1, p. 205.

\bibitem{schep_94} T. J. Schep, F. Pegoraro, and B. N. Kuvshinov, Phys. Plasmas {\bf 1}, 2843 (1994).

\bibitem{simakov_06} A. N. Simakov, L. Chac\'on, and D. A. Knoll, Phys. Plasmas {\bf 13}, 082103 (2006).

%\bibitem{zocco_09} A. Zocco, L. Chac\'on, and A. N. Simakov, Phys. Plasmas {\bf 16}, 110703 (2009).

\end{thebibliography}
\end{document}